\documentclass{article}
\usepackage{spconfa4,amsmath,graphicx}
\usepackage[caption=false,font=footnotesize]{subfig} 
\usepackage{lipsum} 
\usepackage{multirow}
\newcommand{\transp}{^{\textit{\scriptsize{T}}}} 
\renewcommand{\vec}[1]{\boldsymbol{\mathbf{#1}}} 
\usepackage{xcolor} 
\usepackage{cite}

\definecolor{ieeeblue}{RGB}{0, 98, 155}
\definecolor{nlms}{RGB}{0, 181, 226}
\definecolor{pemnlms}{RGB}{119, 37, 131}
\definecolor{neuralAFC}{RGB}{0, 132, 61}
\definecolor{DFC-OL}{RGB}{255, 209, 0}
\definecolor{DFC-IL}{RGB}{186, 12, 47}
\title{In-the-Loop Training of Deep Feedback Cancellation for Hearing Aids}
\name{Svantje Voit and Simon Doclo\thanks{This research was funded by the Deutsche Forschungsgemeinschaft (DFG, German Research Foundation) -- Project ID 352015383 -- SFB 1330 C1, and Germany's Excellence Strategy -- EXC 2177/2 -- Project ID 390895286. Simulations were conducted on the HPC cluster ROSA funded by the DFG under INST 184/225-1 FUGG.}}
\address{Dept. of Medical Physics and Acoustics and Cluster of Excellence Hearing4all,\\ Carl von Ossietzky Universität Oldenburg, Germany, \{svantje.voit, simon.doclo\}@uol.de}
\begin{document}
\ninept
\maketitle
\begin{abstract}
Acoustic feedback limits the maximum gain in hearing aids. In addition to several approaches based on adaptive filtering, recently a deep-neural-network-based feedback cancellation (DFC) approach has been proposed, which is trained via an open-loop framework. Since open-loop-trained DFC (DFC-OL) can become unstable during inference at high gains, in this paper we propose an in-the-loop-trained DFC (DFC-IL) that integrates the DFC directly into the optimisation loop. 
This allows the model to be exposed to unstable conditions during training. A two-stage training strategy involving pre-training on stable systems and fine-tuning on a wider gain range enables DFC-IL to learn robust howling reduction. Experimental results on measured feedback paths demonstrate that in scenarios with small gains, the proposed DFC-IL performs similarly to DFC-OL, and both exceed the performance of adaptive filters. In scenarios with high amplification gains, DFC-IL clearly outperforms DFC-OL by maintaining system stability.
\end{abstract}
\section{Introduction}
\label{sec:intro}
Acoustic feedback in hearing aids occurs due to the coupling between the loudspeaker and the microphone, forming a closed-loop acoustic system \cite{Waterschoot2010}.
Acoustic feedback can significantly impair sound quality and restrict the maximum gain \cite{Waterschoot2010, Spriet2008}. 
If the gain is too high, the system can become unstable, resulting in unpleasant howling artefacts \cite{Waterschoot2010, Kates2002, Zhang2023}. 
Therefore, robust feedback control algorithms are a key part of the sound processing in hearing aids \cite{Waterschoot2010, Spriet2008, Kates2002}.

Various techniques have been developed to mitigate acoustic feedback, including phase modulation \cite{Guo2012phase, Guo2016}, noise injection \cite{Guo2012, Nakagawa2014}, and multi-microphone configurations \cite{Tran2018, Schepker2019, Nakagawa2014mm}, of which acoustic feedback cancellation (AFC) is considered one of the most effective solutions \cite{Waterschoot2010, Strasser2015}. In the AFC framework, the impulse response of the feedback path is typically estimated using an adaptive filter, e.g., the normalised least mean squares (NLMS) algorithm \cite{Haykin2008}. However, the resulting estimate is typically biased due to the inherent correlation between the loudspeaker and microphone signals \cite{Waterschoot2010}.

To reduce this bias, the prediction error method (PEM) is a widely adopted approach in adaptive feedback cancellation \cite{Spriet2008, Spriet2005, Spriet2006}, aiming to decorrelate the signal components by modelling the incoming source signal as an autoregressive process and applying its inverse whitening filter to the adaptive filter inputs. Various enhancements of the PEM framework have been explored, e.g., using the affine projection algorithm \cite{Tran2016}, proportionate NLMS \cite{Tran2017}, and Kalman filtering \cite{Bernardi2015}. However, since PEM cannot perfectly decorrelate the loudspeaker and microphone signals, robust feedback-path estimation using classical adaptive filter algorithms remains challenging.
Consequently, the use of an affine combination \cite{Schepker2016} or a switching mechanism \cite{Nordholm2018} between the NLMS and PEM-NLMS algorithms with distinct step-sizes has been investigated. 

In \cite{Soleimani2023}, the NeuralAFC algorithm was proposed, where a deep neural network (DNN) predicts masks to derive a step-size for adaptive filter control in AFC. The NeuralAFC performs frame-wise filter updates, following previous work on DNN-based adaptation control for echo cancellation \cite{Haubner2022}. A combination of NeuralAFC with the PEM was proposed in \cite{Zhan2025}. A fully DNN-based deep feedback cancellation (DFC) approach was introduced in \cite{Lydaki2025}, where the DNN directly estimates the feedback path from the loudspeaker and microphone signals.
Crucially, in \cite{Lydaki2025} the DFC is trained via an {open-loop training} framework, where the training data is obtained from a {closed-loop acoustic system} without feedback cancellation. 
This restricts the open-loop-trained DFC (DFC-OL) only to be trained for stable closed-loop conditions, i.e., for low gains.
In this paper, we propose to incorporate the DFC directly into the training loop to ensure consistency between training and inference. 
Specifically, the proposed in-the-loop-trained DFC (DFC-IL) combines the DNN-based feedback-path estimation of the DFC from \cite{Lydaki2025} with the frame-wise in-the-loop training inspired by the NeuralAFC framework from \cite{Soleimani2023}. 
This allows the DFC-IL to be exposed to challenging and even unstable scenarios within the {closed-loop acoustic system} during its optimisation. 
As a result, the proposed DFC-IL algorithm provides a higher added stable gain than the DFC-OL. This enables to provide larger gains to hearing aid users while maintaining system stability.

\section{AFC Methods}
\label{sec:afc}
\begin{figure}[!t]
  \centering
  \vspace{-1ex}
  \includegraphics[scale=1]{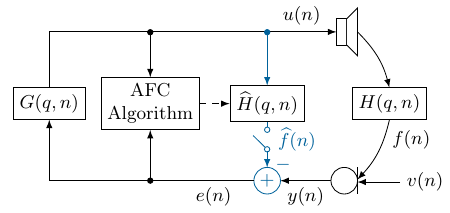}\vspace{-1ex}
  \caption{Processing scheme with acoustic feedback cancellation (AFC). Blue branch indicates AFC processing. Closed switch means in-the-loop processing and open switch means open-loop processing.}
  \label{fig:afc}
  \vspace{-3ex}
\end{figure}
Figure~\ref{fig:afc} illustrates the typical signal processing scheme of a hearing aid. The microphone signal $y(n)$, with $n$ the discrete-time index, consists of the desired incoming source signal $v(n)$ and the undesired feedback signal $f(n)$, i.e.,
\begin{equation}
    y(n) = v(n) + f(n) = v(n) + H(q,n) u(n) \, ,
\end{equation}
where $u(n)$ is the loudspeaker signal and $H(q,n)$ represents the acoustic feedback path, with $q^{-1}$ the discrete-time delay operator. The feedback path is modelled as a finite impulse response filter of length $L_H$, defined as $H(q,n) = \vec{q}_{L_{H}}\transp \vec{h}(n)$, where $\vec{h}(n) = [h_0(n), \dots, h_{L_{H}-1}(n)]\transp$ denotes the filter vector and $\vec{q}_{L_{H}} = [1, q^{-1}, \dots, q^{-L_{H}+1}]\transp$ is the delay vector.

The forward path $G(q,n)$ represents the hearing aid processing. While in general $G(q,n)$ can be time- and frequency-varying, here we assume a fixed forward path $G(q,n) = G\, q^{-d_G}$, characterised by a broadband gain $G$ and a processing delay $d_G \geq 1$.

If no AFC is performed, the loudspeaker signal is
\begin{equation}
    u(n) = G(q,n)\,y(n) \, .
\end{equation}
According to the Nyquist stability criterion \cite{Kates2002, Nyquist1932}, the closed-loop system is stable if the closed-loop gain satisfies
\begin{equation}
    \mathcal{C}(n) = \max_\omega \{ | G(\omega,n) H(\omega,n) | \} < 0 \, \mathrm{dB} \, ,
\end{equation}
where $G(\omega,n)$ and $H(\omega,n)$ denote the frequency responses of the forward and feedback paths, with $\omega$ the radian frequency. If this condition is violated, the system may become unstable, resulting in undesired howling artefacts.

To mitigate this, AFC algorithms employ an adaptive filter $\widehat{H}(q,n) = \vec{q}_{L}\transp \widehat{\vec{h}}(n)$ of length $L$ to estimate the possibly time-varying feedback path and subtract the estimated feedback signal $\widehat{f}(n)= \widehat{H}(q,n) u(n)$ from the microphone signal $y(n)$ \cite{Waterschoot2010, Spriet2008, Kates2002}. This yields the feedback-compensated error signal
\begin{equation}
\label{eq:error_signal}
    e(n) = v(n) + \left(H(q,n) - \widehat{H}(q,n)\right) u(n) \, ,
\end{equation}
which is subsequently processed by the forward path to generate the loudspeaker signal $u(n) = G(q,n) e(n)$. In this configuration, the system stability is governed by the effective closed-loop gain
\begin{equation} \label{eq:eclg}
    \mathcal{C}_E(n) = \max_\omega \{ | G(\omega,n) (H(\omega,n) - \widehat{H}(\omega,n)) | \} \, .
\end{equation}
The system is stable if $\mathcal{C}_E(n) < 0 \, \mathrm{dB}$, implying that a precise estimate $\widehat{H}(q,n)$ allows for higher forward-path gains $G$ without inducing instability.

\section{In-the-Loop Deep Feedback Cancellation}
\label{sec:proposed}
In \cite{Lydaki2025}, the DFC model has been trained using an open-loop framework, where the AFC algorithm is excluded during the data generation phase of the training process. While this enables efficient generation of training data, it limits the dataset to stable acoustic systems where the closed-loop gain satisfies $\mathcal{C}(n) < 0$ dB. This training framework causes a mismatch between DNN training and inference, as the network is never exposed to the effects of residual feedback or instability during its optimisation.

To address this mismatch, we propose an in-the-loop training framework for DFC, inspired by \cite{Soleimani2023} where in-the-loop training was used to train a DNN for step-size control in adaptive feedback cancellation. Within this framework, the filter $\widehat{\vec{h}}(n)$ estimated by the DNN is utilised to compute the feedback signal estimate $\widehat{f}(n) $ and the closed-loop error signal $e(n)$ during the forward pass of the training process. This recursive integration ensures that the DNN parameters are optimised directly with respect to the model's own feedback-cancellation performance. Consequently, the network can be exposed to potentially unstable conditions.

\begin{figure}[!t]
  \centering
  \includegraphics[scale=.8]{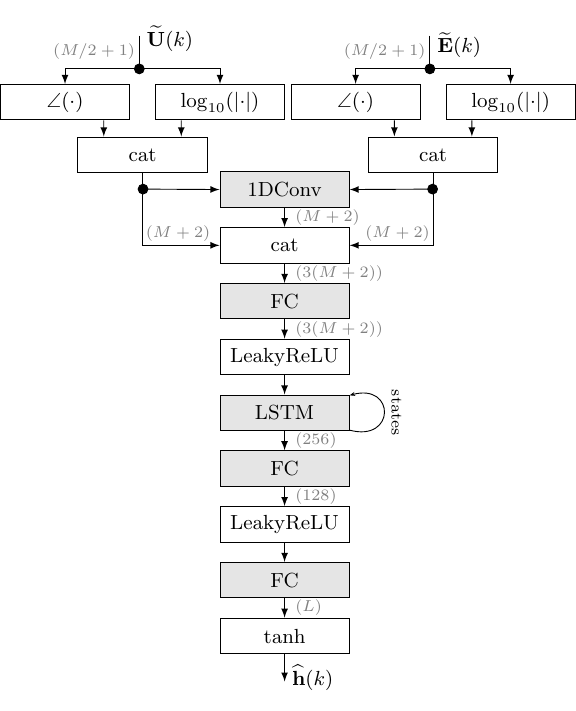}\vspace{-1ex}
  \caption{Block diagram of the in-the-loop-trained DFC architecture with normalised spectra $\widetilde{\vec{U}}(k)$ and $\widetilde{\vec{E}}(k)$ as input and time-domain filter $\widehat{\vec{h}}(k)$ as output. Grey blocks indicate the trainable layers.}
  \label{fig:dfc}
\end{figure}
The considered DNN architecture is based on the DFC architecture proposed in \cite{Lydaki2025}, with specific modifications to enforce temporal consistency and reduce computational complexity via frame-wise updates \cite{Soleimani2023,Proakis1996}. 
The DNN operates in the frequency domain to effectively extract spectral correlations. The time-domain frames $\vec{u}(k) = [u(kR-M+1), \dots, u(kR)]\transp$ and $\vec{e}(k) = [e(kR-M+1), \dots, e(kR)]\transp$, with frame length $M$, frame shift $R$ and frame index $k$, are transformed via an $M$-point discrete Fourier transform to yield the spectra $\vec{U}(k)$ and $\vec{E}(k)$. These are normalised following \cite{Lydaki2025} to obtain $\widetilde{\vec{U}}(k) = \vec{U}(k) / \lVert\vec{U}(k)\rVert_2$ and $\widetilde{\vec{E}}(k) = \vec{E}(k) / \lVert\vec{U}(k)\rVert_2$.
As depicted in Figure~\ref{fig:dfc}, the normalised spectra are concatenated as two input channels and processed by a 1D convolutional layer with kernel size 5 to produce a single-channel spectral envelope. This learned feature is again concatenated with the normalised spectra $\widetilde{\vec{U}}(k)$ and $\widetilde{\vec{E}}(k)$, spanning a $(3(M+2))$-dimensional feature space. Following the DFC architecture, these features are subsequently processed by a fully connected (FC) layer with LeakyReLU activation, an LSTM layer with 256 units, and an additional FC layer with LeakyReLU. Finally, an FC layer with $\tanh$ activation provides the time-domain filter coefficients $\vec{\widehat{h}}(k) = [\widehat{h}_{1}(k), \dots, \widehat{h}_{L}(k)]\transp$ as
\begin{equation}
    \vec{\widehat{h}}(k) = \text{DNN}\left\{ \widetilde{\vec{U}}(k), \widetilde{\vec{E}}(k) \right\} \, .
\end{equation}

Unlike in \cite{Lydaki2025}, we omit average pooling and recursive smoothing, thereby strictly limiting the temporal context to a single frame.
To ensure stable convergence, a two-stage training strategy is implemented using the Adam optimiser \cite{Kingma2015} in its standard configuration. 
In the first stage, the network is pre-trained on stable systems with a closed-loop gain range of $\mathcal{C}(n) \in [-6, 0)$~dB, using a batch size of 32 utterances and a learning rate of $10^{-3}$. In the second stage, the model is fine-tuned using a reduced learning rate of $10^{-4}$ across an expanded gain range of $\mathcal{C}(n) \in [-6, 16]$~dB exposing the model to unstable scenarios.
The loss function to train the network is the normalised Euclidean system distance \cite{Lydaki2025, Soleimani2023}
\begin{equation}
    \text{NESD}_{\text{dB}}(\vec{h}(k),\widehat{\vec{h}}(k)) = 20 \log_{10} \left( \frac{{\lVert \vec{h}(k) - \widehat{\vec{h}}(k) \rVert}_2}{{\lVert\vec{h}(k)\rVert}_2 } \right) 
\end{equation}
averaged across all frames and utterances within the batch. 

\begin{table*}[ht]
\centering
\caption{Averaged ECLG across all test utterances and all 15 seconds, including convergence speed and steady-state performance. Negative numbers indicate stability, while positive numbers indicate potential instability.}\vspace{.2ex}
\begin{tabular}{c||c|c|c|c|c}
\hline 
\multirow{2}{*}{\bf Method}& \multicolumn{5}{c}{\bf Scenario} \\[.2ex] \cline{2-6}
& $\mathcal{C}_{1}(n) \in {[-5, 0)} $ dB & $\mathcal{C}_{2}(n) \in {[0, 5]} $ dB & $\mathcal{C}_{3}(n) \in {[5, 10]} $ dB & $\mathcal{C}_{4}(n) \in {[10, 15]} $ dB & Average \\\hline \hline 
NLMS      & $-3.0$ & $-2.0$ & $-0.2$ & $3.0$ & $-0.2$ \\ 
PEM-NLMS \cite{Spriet2008}   & $-6.0$ & $-4.3$ & $-1.6$ & $2.2$ & $-1.9$ \\ 
NeuralAFC \cite{Soleimani2023}  & $-6.9$ & $-5.9$ & $-3.4$ & $1.1$ & $-3.2$ \\ 
DFC-OL \cite{Lydaki2025}      & $\boldsymbol{-12.2}$ & $\boldsymbol{-8.9}$ & $-3.2$ & $4.1$ & $-2.1$ \\ 
\bf DFC-IL  (proposed)    & $-9.9$ & $-6.8$ & $\boldsymbol{-3.5}$ & $\boldsymbol{-0.7}$ & $\boldsymbol{-4.6}$ \\ \hline
\end{tabular}
\label{tab:results}
\end{table*}
\section{Experimental Setup and Results}
\label{sec:results}
\subsection{Training and Test Data}
\label{subsec:data}
For training and evaluation, we utilise a diverse dataset comprising a total of $15,\!160$ utterances at a sampling frequency of 16\,kHz, combining speech signals $v(n)$ with both measured and synthetic feedback paths $\vec{h}(n)$. The measured feedback paths include $5,\!740$ impulse responses from the Hearpiece database \cite{Denk2021}, covering open- and closed-fittings, various loudspeaker and microphone placements, and multiple persons. Additionally, 420 measured feedback paths from a two-microphone behind-the-ear hearing aid, recorded on a dummy head \cite{Sankowsky2015}, are included to account for different ear canal geometries and venting conditions. To further increase diversity, we added $9,\!000$ synthetic open-fit feedback paths generated according to the model in \cite{Zheng2022}. All feedback paths are truncated to a length of $L_H=64$. As suggested in \cite{Lydaki2025}, to simulate realistic conditions, the magnitude responses $\lvert H(\omega,n)\rvert$ are normalised, such that their maxima follows a uniform distribution between $-20$\,dB and $-10$\,dB.

The incoming source signals are taken from the LibriSpeech \cite{librispeech} corpus, utilising 80 speakers for training and validation and 6 distinct speakers for testing. Each feedback path is combined with a random speech signal. The feedback-path data is partitioned as follows:
\begin{itemize}
    \item \textbf{Training:} $12,\!928$ paths (4,928 measured, $8,\!000$ synthetic);
    \item \textbf{Validation:} $1,\!616$ paths (616 measured, $1,\!000$ synthetic);
    \item \textbf{Test:} 616 paths (exclusively measured).
\end{itemize}
The test set relies solely on measured feedback paths to ensure a realistic performance evaluation. 

The forward-path delay is fixed at $d_G=160$ samples, corresponding to 10\,ms. Training and validation utterances are two seconds long, while the test utterances are 15 seconds long and include an abrupt feedback path switch at 7.5 seconds to evaluate tracking performance.

\subsection{Algorithms}
\label{subsec:methods}
We evaluate the performance of the proposed DFC-IL model with a range of classical and state-of-the-art neural feedback cancellation algorithms. For all algorithms the filter length is set to $L=64$, matching the feedback-path length $L_H$.
\subsubsection{Classical Baselines}
As classical AFC algorithms, we include the NLMS algorithm with a step-size of $\mu=0.01$, and the PEM-NLMS algorithm \cite{Spriet2008} using a pre-filter of length 20 estimated via Levinson-Durbin recursion \cite{Waterschoot2010} and a step size of $\mu=0.002$.
\subsubsection{Neural Baselines}
As the first neural baseline, we consider {NeuralAFC} \cite{Soleimani2023}, where a DNN predicts a frame-wise step-size to control an adaptive filter. It uses a frame length of $M=128$ and a frame shift of $R=63$ samples. Consistent with \cite{Soleimani2023}, it is trained in-the-loop on 10-second utterances with a random feedback-path switch between 2 and 8 seconds and closed-loop gains $\mathcal{C}(n) \in [-5, 5]$~dB.

As second neural baseline, we consider the DFC model proposed in \cite{Lydaki2025}, denoted as {DFC-OL}. This model is trained exclusively on stable systems, with $\mathcal{C}(n) \in [-6, 0)$~dB. It is trained on 2-second utterances using $M=128$ and $R=32$. It should be noted that during inference, DFC-OL utilises a significantly larger temporal context with $M=1696$ and $R=1$ as described in \cite{Lydaki2025}.
\subsubsection{Proposed DFC-IL}
The proposed DFC-IL model is trained in-the-loop on 2-second utterances using $M=128$ and $R=63$. As described in Section~\ref{sec:proposed}, it undergoes a two-stage training process: initial pre-training on stable systems with $\mathcal{C}(n) \in [-6, 0)$~dB, followed by fine-tuning on potentially unstable scenarios of up to $\mathcal{C}(n) = 16$~dB.

\subsection{Evaluation Metrics}
\label{subsec:metric}
The performance of the algorithms is evaluated using two metrics. 
To assess the overall system stability, we utilise the effective closed-loop gain (ECLG) defined in \eqref{eq:eclg}.
When the ECLG exceeds 0 dB, the system can become unstable and produce unpleasant howling \cite{Strasser2015}.
In addition, we utilise the added stable gain (ASG), representing the additional gain that can be provided to a user, i.e.,
\begin{equation} \label{eq:asg}
    \text{ASG}(n) =  \frac{\max_\omega \{\lvert H(\omega,n) \rvert\} }{\max_\omega \{ \lvert H(\omega,n) - \widehat{H}(\omega,n)\rvert\} } \, .
\end{equation}
\begin{figure}[t!]
  \centering
  \subfloat[Scenarios with stable closed-loop systems, $\mathcal{C}_{1}(n) \in [-5, 0) $ dB]{\includegraphics[scale=1]{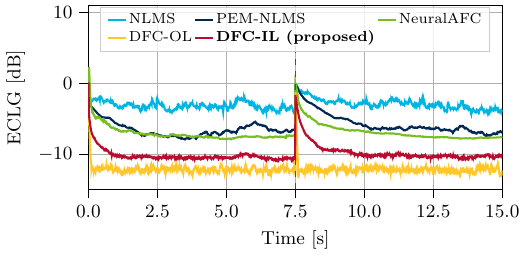}%
  \label{fig:G1_ECLG}}

  \subfloat[Scenarios with unstable closed-loop systems, $\mathcal{C}_{4}(n) \in {[10, 15]} $ dB]{\includegraphics[scale=1]{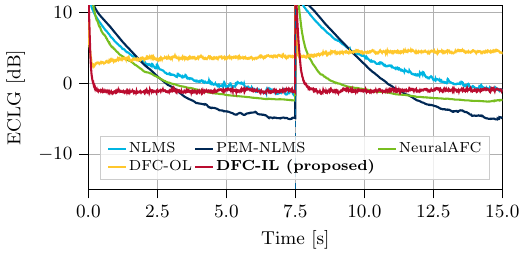}%
  \label{fig:G4_ECLG}}
  \caption{Effective closed-loop gain for different scenarios with a feedback-path change at 7.5\,s.}
  \label{fig:ECLG_results}
\end{figure}
\begin{figure}[t!]
  \centering \vspace{0.55ex}
  \subfloat[Scenarios with stable closed-loop systems, $\mathcal{C}_{1}(n) \in [-5, 0) $ dB]{\includegraphics[scale=1]{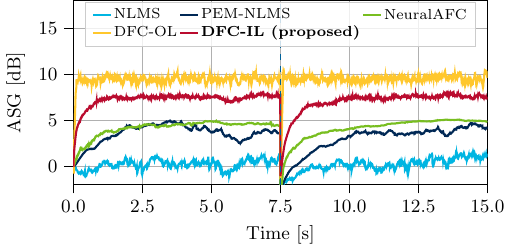}%
  \label{fig:G1_ASG}}

   \vspace{0.6ex}
  \subfloat[Scenarios with unstable closed-loop systems, $\mathcal{C}_{4}(n) \in {[10, 15]} $ dB]{\includegraphics[scale=1]{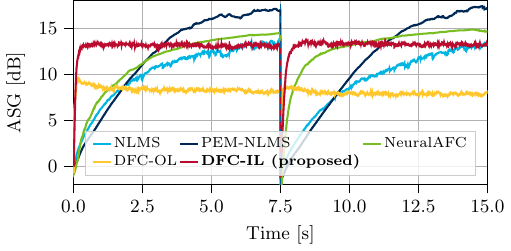}%
  \label{fig:G4_ASG}}
  \caption{Added stable gain for different scenarios with a feedback-path change at 7.5\,s.}
  \label{fig:ASG_results}
\end{figure}
\subsection{Experimental Results and Analysis}
Evaluation is conducted for four scenarios with increasing closed-loop gain ranges: $\mathcal{C}_1(n) \in [-5, 0)$~dB, $\mathcal{C}_2(n) \in [0, 5]$~dB, $\mathcal{C}_3(n) \in [5, 10]$~dB, and $\mathcal{C}_4(n) \in [10, 15]$~dB. For all considered algorithms, Table~\ref{tab:results} presents the ECLG averaged across the test utterances.

First, it can be observed for all scenarios that PEM-NLMS achieves a lower ECLG than NLMS, illustrating the effectiveness of its decorrelation mechanism. Second, it can be observed for all scenarios, that NeuralAFC achieves a lower ECLG than both classical adaptive filtering algorithms, even in high-gain scenarios ($\mathcal{C}_3$ and $\mathcal{C}_4$) for which it was not explicitly trained. Finally, it can be observed that for low-gain scenarios ($\mathcal{C}_1$ and $\mathcal{C}_2$), DFC-OL achieves the lowest ECLG of all algorithms, whereas for high-gain scenarios ($\mathcal{C}_3$ and $\mathcal{C}_4$), the proposed DFC-IL achieves the lowest of all algorithms. On average over all scenarios, the proposed DFC-IL outperforms all other considered algorithms in terms of ECLG.

To evaluate tracking and convergence, Figures~\ref{fig:ECLG_results} and~\ref{fig:ASG_results} show the time-dependent effective closed-loop gain and the added stable gain for the (stable) low-gain scenario $\mathcal{C}_1$ and the high-gain scenario $\mathcal{C}_4$. Each curve represents the averaged performance across the test utterances for all considered algorithms.
First, it can be observed for both scenarios that the NLMS and PEM-NLMS adaptive algorithms converge slower than the neural feedback cancellation algorithms, especially after the feedback path change at 7.5\,s. Nevertheless, it should be noted that in the high-gain scenario the PEM-NLMS algorithm converges to the lowest ECLG and highest ASG of all algorithms. Second, it can be observed that for both scenarios the NeuralAFC algorithm converges slower than DFC-OL and DFC-IL. For the low-gain scenario, the ECLG and ASG values obtained by the NeuralAFC algorithm at convergence are worse than for DFC-OL and DFC-IL.

On the other hand, for the high-gain scenario the ECLG and ASG values obtained by the NeuralAFC algorithm at convergence are similar to DFC-IL but better than DFC-OL. Finally, comparing the DFC-OL and DFC-IL algorithms, it can be observed that for the low-gain scenario DFC-OL outperforms DFC-IL both in convergence speed as well as ECLG and ASG values at convergence, whereas for the high-gain scenario the convergence speed of DFC-OL and DFC-IL is similar but the ECLG and ASG values at convergence are much better for DFC-IL than DFC-OL. This confirms the findings from Table~\ref{tab:results}, showing that DFC-OL outperforms DFC-IL for matched conditions but does not generalise well for large closed-loop gains. Overall, the best overall performance for both scenarios in terms of convergence speed and ECLG and ASG values at convergence is obtained by the proposed DFC-IL.

\section{Conclusion}
\label{sec:conclusion}
In this paper, we introduced an in-the-loop training framework for deep feedback cancellation that aligns the optimisation process with inference and allows unstable conditions to be included during training. By incorporating the feedback cancellation algorithm into the training loop, it was shown that in-the-loop-trained DFC demonstrates robust performance across a wide range of gains, while maintaining the fast tracking ability of open-loop-trained DFC. 

\bibliographystyle{IEEEbib}
\bibliography{refs}

\end{document}